\documentclass[conference]{IEEEtran}
\IEEEoverridecommandlockouts
\usepackage{cite}
\usepackage{amsmath,amssymb,amsfonts}
\usepackage{algorithmic}
\usepackage{graphicx}
\usepackage{textcomp}
\usepackage{xcolor}
\usepackage{color}
\usepackage{soul}

\usepackage[hyphens,spaces,obeyspaces]{url}

\newcommand{\covid}[1]{COVID-19\ }

\newcommand\blfootnote[1]{%
  \begingroup
  \renewcommand\thefootnote{}\footnote{#1}%
  \addtocounter{footnote}{-1}%
  \endgroup
}

\usepackage{booktabs}
\def\BibTeX{{\rm B\kern-.05em{\sc i\kern-.025em b}\kern-.08em
    T\kern-.1667em\lower.7ex\hbox{E}\kern-.125emX}}
\begin{document}

\title{The Role of the Crowd in Countering Misinformation: A Case Study of the \\COVID-19 Infodemic}

\author{Nicholas Micallef\textsuperscript{*}~\textsuperscript{1}, Bing He\textsuperscript{*}~\textsuperscript{2}, Srijan Kumar~\textsuperscript{2}, Mustaque Ahamad~\textsuperscript{2}, Nasir Memon~\textsuperscript{3} \\~\textsuperscript{1} New York University Abu Dhabi,~\textsuperscript{2} Georgia Institute of Technology,~\textsuperscript{3} New York University \\~\textsuperscript{1}\texttt{nicholas.micallef@nyu.edu},~\textsuperscript{2}\texttt{\{bhe46,srijan,mustaq\}@gatech.edu},~\textsuperscript{3} \texttt{memon@nyu.edu}
\\}

\IEEEoverridecommandlockouts
\IEEEpubid{\makebox[\columnwidth]{
978-1-7281-6251-5/20/\$31.00~\copyright2020 IEEE
\hfill} \hspace{\columnsep}\makebox[\columnwidth]{ }}

\maketitle

\IEEEpubidadjcol

\blfootnote{* co-first authors. The first two authors contributed equally to this work.}

\begin{abstract}

Fact checking by professionals is viewed as a vital defense in the fight against misinformation. 
While fact checking is important and its impact has been significant, fact checks could have limited visibility and may not reach the intended audience, such as those deeply embedded in polarized communities. Concerned citizens (i.e., the crowd), who are users of the platforms where misinformation appears, can play a crucial role in disseminating fact-checking information and in countering the spread of misinformation. To explore if this is the case, we conduct a data-driven study of misinformation on the Twitter platform, focusing on tweets related to the \covid\ pandemic, analyzing the spread of misinformation, professional fact checks, and the crowds response to popular misleading claims about COVID-19.

In this work, we curate a dataset of false claims and statements that seek to challenge or refute them. We train a classifier to create a novel dataset of 155,468 COVID-19-related tweets, containing 33,237 false claims and 33,413 refuting arguments. Our findings show that professional fact-checking tweets have limited volume and reach. In contrast, we observe that the surge in misinformation tweets results in a quick response and a corresponding increase in tweets that refute such misinformation. More importantly, we find contrasting differences in the way the crowd refutes tweets, some tweets appear to be opinions, while others contain concrete evidence, such as a link to a reputed source. Our work provides insights into how misinformation is organically countered in social platforms by some of their users and the role they play in amplifying professional fact checks. These insights could lead to development of tools and mechanisms that can empower concerned citizens in combating misinformation. The code and data can be found in this link.\footnote{\url{http://claws.cc.gatech.edu/covid_counter_misinformation.html}}

\end{abstract}

\begin{IEEEkeywords}
Misinformation, Counter-misinformation, Social Media, Dataset
\end{IEEEkeywords}

\section{Introduction}

The world-wide spread of \covid\ has led to considerable amount of related misinformation on the web and the social media ecosystem. 
WHO has termed the situation as a global  infodemic~\cite{who2020novel}.
As social media platforms become a primary means to acquire and exchange news and information during crisis times such as \covid\~\cite{mendoza2010twitter,viewing2010microblogging,burnap2014tweeting,starbird2014rumors,oyeyemi2014ebola,miller2017what,ortiz2017yellow}, the lack of clear distinction between true and false information can be dangerous. 
Some of the false claims related to \covid\ have already had severe harmful consequences, including violence\cite{bbc2020novel} and over 800 deaths~\cite{islam2020covid-19}. 
Thus, combating the spread of false information is of critical importance.

Professional fact checkers can play an important role in controlling the spread of misinformation on online platforms~\cite{wintersieck2017debating}. 
During the \covid\ infodemic, the International Fact Checking Network (IFCN) verified over 6,800 false claims related to the pandemic until May 20, 2020. 
Social media platforms use these fact checks to flag and sometimes remove misinformation content.
However, false information still prevails on social platforms because the ability of fact checking organizations to use social media to disseminate their work can be limited~\cite{shahi2020exploratory}. For example, on Facebook, content from the top 10 websites spreading health misinformation had almost four times as many estimated views as equivalent content from reputable organizations (e.g., CDC, WHO).\footnote{\url{https://secure.avaaz.org/campaign/en/facebook_threat_health/}}

In addition to professional fact checkers, ordinary citizens, who are concerned about misinformation, can play a crucial role in organically curbing its spread and impact. Compared to professional fact checkers, concerned citizens, who are users of the platform where misinformation appears, have the ability to directly engage with people who propagate false claims either because of ignorance or for a malicious purpose. 
They can back up their arguments using professional fact checks and trusted sources, whenever available. 
The cohort of ordinary citizens is also commonly referred to as \textit{crowd}.
Thus, the role of crowd or citizens who are concerned about misinformation can be critically important. The goal of this work is to study the nature and extent of the role that concerned citizens play in responding to misinformation.  


We use a broad definition of \textit{misinformation} which includes falsehoods, inaccuracies, rumors, decontextualized truths, or misleading leaps of logic, all regardless of the intention of the spreader~\cite{lazer2018thescience,kumar2018false}. In this work, we focus on \covid\ related misinformation on Twitter and utilize a data-driven approach to investigate how fact checks and other organic user responses attempt to refute and counter it.
We explore two popular misinformation topics: \emph{fake cures} and \emph{5G conspiracy theories}~\cite{brennen2020types}. Fake cures for \covid\ included drinking water, eating garlic or ginger, and salt, and 5G conspiracy theories state that 5G technology is responsible for the spread of \covid\ or that \covid\ does not exist and people are getting sick due to 5G radiations. 

A study of fact checks and concerned citizen responses to misinformation presents several challenges. These include collection of data relevant to misinformation topics and development of a classifier for automated detection of tweets that contain misinformation or responses that refute them. Analysis of tweets that counter misinformation is also necessary to gain insights into how misinformation is refuted. We address these questions and make the following contributions.


\begin{itemize}
    \item {We create a  dataset\footnote{\url{http://claws.cc.gatech.edu/covid_counter_misinformation.html}} of tweets related to misinformation and responses that counter it, by first collecting 6,840 fact checking stories from 95 fact checking websites. We focused on two most popular \covid\  topics and collected 155,468 tweets from 105,772 users over a period of 4 months.} 
    \item We develop a text-based classifier for tweets that represent misinformation and their rebuttals. 
    To seed the classifier, two researchers hand-labeled 4,800 tweets into misinformation, rebuttal and other categories. The classifier created from the labeled data achieves an F1 score that is between 0.7 and 0.8, identifying 33,237 misinformation claims, 33,413 refutations, and 87,377 irrelevant tweets. 
    \item Our analysis of the dataset revealed two types of users responsible for refuting misinformation tweets: professional fact-checkers, i.e., people or organizations who systematically fact check a claim, and concerned citizens,  i.e., the crowd that organically counters false claims. 
    We find that 96\% of refuting tweets are generated by concerned citizens, which highlights their importance.  
    \item Our study reveals that professional fact-checks are not only low in tweet volume, but also receive a significantly lower number of retweets. Also, we find that 67\% of tweets that counter misinformation do not contain URL-based evidence such as fact check websites.  
    \item Our analysis of the methods used for countering misinformation reveals two broad classes. One class of tweets is evidence-based, which includes links to fact checks or other trusted sources. The other class is opinion-based and does not include verifiable evidence. We find that opinion based tweets include more assertive words, use more negative words, are more abusive, and exhibit negative emotions and anger.
    \item Finally, we find that citizen responses come from older and more reputed accounts, while misinformation tends to be spread by more recently created accounts. This could potentially be due to the orchestrated nature of some misinformation campaigns.
    \item Reproducibility: Our code and datasets are available for download at \url{http://claws.cc.gatech.edu/covid_counter_misinformation.html}.
    \end{itemize}

Overall, our findings show that countering of misinformation cannot be studied by considering only fact checks. Our work reveals interesting insights and can potentially lead to a better understanding of how best to leverage the organic countering of misinformation to support and amplify the results of work done by fact checkers. It could also lead to development of tools and mechanisms that can empower concerned citizens to combat misinformation.

\section{Related work}

\begin{table*}[htbp]
    \centering
    \caption{A summary of related work on misinformation and how this work differs from it. 
    \label{tab:sum-related-works}
    }
    \begin{tabular}{|l|p{1.6cm}|p{2cm}|p{2cm}|l|l|p{1cm}|l|}
    \hline
  & Professional fact-checking & Evidence-based citizen response & Opinion-based citizen response  & COVID-19 & Classification & Large-scale analysis & Dataset\\ \hline
 
~\cite{wang2020weak, wang2018eann}
 & & & & & \checkmark & & Wechat, Twitter, Weibo \\ \hline
~\cite{young2020detecting}
        &   &    & \checkmark       & \checkmark& \checkmark   & &   Twitter  \\ \hline
~\cite{friggeri2014rumor}
           &  \checkmark  &  &                &       &               & \checkmark & Facebook  \\ \hline
~\cite{vosoughi2018spread}
       &   &    &  \checkmark        &       &               & \checkmark &   Twitter  \\ \hline
~\cite{mendoza2010twitter, starbird2014rumors}
       & \checkmark  &   &  \checkmark    &       &               &  \checkmark &  Twitter  \\ \hline
~\cite{vraga2017using, van2020seeking}
      &        &  \checkmark  &  \checkmark   &       &               &  &   Twitter   \\ \hline
~\cite{dharawat2020drink}
          &     &   &  \checkmark       & \checkmark  &   \checkmark  & \checkmark &    Twitter  \\ \hline
\hline
This work
    & \checkmark   & \checkmark &  \checkmark   & \checkmark &   \checkmark  & \checkmark &  Twitter  \\ \hline
    \end{tabular}

\end{table*}


Our work builds upon a vast collection of prior work related to the broad field of misinformation and how it is countered. Recent research has focused on \covid\ misinformation, since one of the consequences of this pandemic was the rise of a global infodemic of misinformation. Table~\ref{tab:sum-related-works} presents a comprehensive comparison of our work to previous work.

\subsection{Misinformation and Counter-Misinformation}
Social media contains many types of misinformation, such as  rumors~\cite{friggeri2014rumor}, hoaxes~\cite{kumar2016disinformation}, false news~\cite{vosoughi2018spread} and false information~\cite{kumar2018false}. 
Due to its harmful impact on society, research has explored, analyzed and characterized misinformation and how to counter it~\cite{hassan2019introduction}. These efforts can be grouped into three:\\
\textit{(1) Misinformation or Counter-Misinformation Detection:} Different data sources and methods have been developed to detect misinformation, including user reports through reinforcement learning~\cite{wang2020weak}, 
textual and visual features from posts by adversarial neural network~\cite{wang2018eann}, new user and tweets features with Bayes Network, k-Nearest Neighbors and Adaptive Boosting supervised learning framework~\cite{antoniadis2015model}, 
among others. However, these methods do not study counter-misinformation. On the other hand, some computational methods are designed to limit the propagation of misinformation \cite{litou2017efficient}. However,
they mainly focus on finding a subset of users who initiate the propagation of credible information to limit the spread of misinformation and they do not consider the textual information of the tweet or conduct a downstream analysis.\\
\textit{(2) User Survey-based Experimental Studies:} In ~\cite{vraga2017using, van2020seeking, bode2020right} user surveys are conducted to determine how we can combat misinformation by posting reliable rebuttals on social media platforms. But these studies are small-scale and do not detect or analyze any misinformation.\\
\textit{(3) Observational Studies:} Past research ~\cite{friggeri2014rumor} used comments on rumors that included a URL to fact-checking websites to study the spread of misinformation and counter-misinformation. However, as we show, very few posts receive such comments, while most countering is done without using URLs. Thus, this approach is limited in scale.
Further, Vosoughi et al.~\cite{vosoughi2018spread} used six fact-checking websites to determine whether a tweet containing news information is true or false by text comparison and analyzed their spread. However, fact-checks were not examined. 
Additionally, some works~\cite{bhuiyan2020investigating, roitero2020covid} studied the assessment differences when different people judge the credibility of the articles.
Meanwhile, other works~\cite{mendoza2010twitter, starbird2014rumors} studied the spreading patterns of tweets that contain false rumors and confirmed news, using URLs and news content.
Different from previous work our research investigates the broad information ecosystem which consists of misinformation and its rebuttals. These rebuttals can come from professional fact-checkers or concerned citizens, through opinion-based or evidence-based responses.




\subsection{\covid\ Misinformation Studies}
In recent months, research has analyzed \covid\ misinformation~\cite{cinelli2020covid,kouzy2020coronavirus,shahi2020exploratory}. 
Particularly, Brennen et al.~\cite{brennen2020types} analyzed different sources and claims of \covid\ misinformation and found that most misinformation appears on social networks.  Additionally, other work~\cite{sharma2020covid, huang2020disinformation} used links to low-credibility sources to identify misinformation content on social networks and characterize it by sentiment and tweeting network, while Young et al.~\cite{young2020detecting} identified misinformation by tweet text. 
Furthermore, research used seven different labels to comprehensively label a tweet from different perspectives to better detect and fight misinformation~\cite{alam2020fighting}. Even more, Memon et al.~\cite{memon2020characterizinga} manually labeled misinformation tweets according to fifteen categories and provided a novel misinformation tweet dataset for misinformation community analysis.
However, these works mainly ignore the role of citizen responses and professional fact-checkers in combating \covid\ misinformation and in our work we fill this gap.


The research most relevant to our work is Dharawat et al.~\cite{dharawat2020drink}. They conducted text-based tweet classification and downstream analysis. However, the main difference is that they fail to consider professional fact-checking and evidence-based citizen responses in their analysis. 
Additionally, they only focus on health-related misinformation, while we also cover conspiracy theories to have a broad overview of \covid\ misinformation. 
Another, limitation of previous work is that they mostly focus on  tweets that have links to low-credibility or mainstream sources. This prevents them from analyzing opinion-based tweets, which in our dataset consist of 67\% of the tweets. Hence, to address these shortcomings, in this work, we develop a text-based classifier that allows us to investigate the dataset of misinformation, countering-misinformation and professional fact-checkers tweets.

\section{Data Collection}
We collected data on two of the most popular \covid\ misinformation topics: \emph{fake cures} and \emph{5G conspiracy theories}~\cite{brennen2020types}. Below we describe our process of selecting these topics and finding misinformation and counter-misinformation tweets on these topics.

\subsection{IFCN Dataset}
To have a verified and complete list of COVID-19-related misconception statements, we built a misconception dataset by leveraging the work of fact checkers. Specifically, first, we extracted 6,840 false statements fact-checked by IFCN's CoronaVirusFacts/DatosCoronaVirus alliance.\footnote{\url{https://www.poynter.org/ifcn-covid-19-misinformation/}} 
From these statements, we selected English statements related to two widely fact-checked misinformation topics~\cite{brennen2020types}: (1) \textit{Fake cures} that include drinking water, eating garlic or ginger, and salt. e.g., ``the coronavirus can be cured with a bowl water with ginger''; (2) \textit{5G conspiracy theories} that state 5G technology is responsible for the spread of \covid\ or that \covid\ does not exist and people are getting sick due to 5G radiations.
Our final IFCN dataset consisted of 57 claims associated with fake cures and 32 claims related to 5G.



\subsection{\covid\ Tweet Dataset}
Similar to~\cite{chen2020tracking}, we utilized a keyword-based method to collect COVID-19-related tweets. After generating the IFCN dataset, we identify a collection of keywords related to the entities mentioned in the data. These keywords belong to two sets:

\noindent $\bullet$ \textit{\covid\ keywords}: this set includes widely-used terms associated with COVID-19, such as: COVID-19, covid, corona virus, coronavirus~\cite{chen2020tracking}. \\
\noindent $\bullet$ \textit{Misconception entity keywords}: 5G, drinking water, ginger, garlic and salt. This set was extracted from widely mentioned terms in IFCN misconception statements related to our selected topics. 


Next, we created a \covid\ tweet dataset by combining data from multiple sources. 
First, we retrieved tweets from a publicly available dataset~\cite{chen2020tracking}, which uses Twitter's Streaming API and selected tweets containing the aforementioned keywords. 
Since the streaming API only collects 1\% of the tweets,\footnote{\url{https://developer.twitter.com/en/docs/labs/sampled-stream/overview}} we used Twitter Search API\footnote{\url{https://developer.twitter.com/en/docs/twitter-api/v1/tweets/search/api-reference/get-search-tweets}} to augment our dataset by querying the aforementioned keywords. 

Together, these data sources returned around 8 million tweets spanning 4 months from January 21, 2020 to May 20, 2020. Similarly to other research, to simplify our analysis, we filter out non-English tweets~\cite{hossain2020detecting} and retweets. Finally, we get 155,468 tweets from 105,772 users for the selected two topics.

\subsection{Professional Fact-Checking Tweet Dataset}
In addition to the above tweets, we also collected tweets from fact-checking organizations' Twitter accounts. We manually compiled a list of Twitter handles of the organizations mentioned in the IFCN dataset. Then we used Python's Tweepy API\footnote{\url{https://www.tweepy.org/}} to extract tweets posted between January 21, 2020 and May 20, 2020. Later, we manually filtered these tweets by keeping only those tweets that contained an explicit link to one of the 89 claims listed in the IFCN dataset. This data served as our professional fact-checking tweet dataset.

\section{Hand-labeled Annotation}

Traditional methods usually use links to low-credibility sources to label tweets as misinformation~\cite{sharma2020covid}. However, in our \covid\ tweet dataset, only around 10\% of the tweets have links. Since there are no ground-truth labels for the remaining 90\%, we manually label a subset of tweets based on tweet content and create a textual classifier to label the entire dataset. 

\subsection{Label Definition}
We provide the definitions of the labels to make the annotation more clear.
Based on previous misinformation research~\cite{lazer2018thescience,kumar2018false}, we define three labels:

\noindent $\bullet$ \textbf{Misinformation Tweets}: Since in this research we are interested in identifying a broader set of \covid\ misinformation tweets, we consider misinformation tweets to include falsehoods, inaccuracies, rumors, decontextualized truths, or misleading leaps of logic, all regardless of the intention of the spreader. See Tweets $1~\&~2$ in Table~\ref{tab:tweetannotation}.

\noindent $\bullet$ \textbf{Counter-Misinformation Tweets}: Tweets in this category refute false claims and can be further categorized in one of the following two groups:. 
\begin{itemize}
\item[--] \textbf{Professional Fact-check Tweets}: These counter-misinformation tweets are generated by professional fact-checkers and were extracted from the professional fact-checking tweet dataset. See Tweet 3 in Table~\ref{tab:tweetannotation}.
\item[--] \textbf{Concerned Citizen Tweets}: Tweets that counter, question, or refute misinformation and are not generated by a professional fact-checking organization. See Tweets $4,~5~\&~6$ in Table~\ref{tab:tweetannotation}. In the analysis we further categorize these tweets into: (a) \emph{Opinion-based citizen responses} and (b) \emph{Evidence-based citizen responses}. The aim is to distinguish between tweets that do not use any checkable evidence and those that link to some evidence.
\end{itemize}



\noindent $\bullet$ \textbf{Irrelevant Tweets}: Any tweet that does not include misinformation or counter-misinformation. The content may still discuss \covid\ . See tweet 7 in Table~\ref{tab:tweetannotation}. 


\subsection{Annotation Process}
The annotation task aims to generate ground-truth for tweets by categorizing them as misinformation, counter-misinformation, or irrelevant. The set of professional fact-checking tweets is deterministic and rule-based, so it does not require hand-labeling or building a classifier. 

To hand-label tweets, two authors annotated 4,800 tweets (specifically, 2,400 tweets for each topic with 800 tweets for misinformation tweets, counter-misinformation tweets, and irrelevant tweets). To assess agreement levels, a random sample of 300 tweets was annotated by both annotators. The inter-rater agreement measured by Kappa score~\cite{schuster2004note} between the two annotators was 0.782, which shows substantial agreement.


\textit{Annotation process:} This was a time consuming and difficult process because social media data is noisy and this process can be subjective. Hence, we conducted this process over several iterations. After each iteration a meeting was conducted to discuss dubious tweets, following the setup used by other research~\cite{zubiaga2015towards,alam2020fighting}. 
Specifically, to annotate the tweets we use the following process. We start by asking the question: \emph{Does this tweet refer to a misconception statement that was verified by IFCN's fact-checking alliance?} If the answer is yes, we determine whether the tweet is supporting the misconception claim or countering it. For instance, Tweet 1 in Table~\ref{tab:tweetannotation} supports the verified false claim which states that garlic could prevent COVID-19, so we annotate it as misinformation tweet. When a tweet counters a false claim, such as when people categorically state that garlic has no effect on coronavirus (see Tweet 4 in Table~\ref{tab:tweetannotation}), we annotate it as a counter-misinformation tweet. If the tweet is neither misinformation nor counter-misinformation, such as Tweet 7 in Table~\ref{tab:tweetannotation}, which is just talking about the surging increase in prices of garlic, we annotate the tweet as an irrelevant tweet. 

Labeling was not always straightforward. For instance, there were tweets which did not refer to any of the false claims that were verified by IFCN. We did not discard these tweets because they could still spread or counter misinformation~\cite{qazvinian2011rumor}.
If those tweets included falsehoods, inaccuracies, rumors, decontextualized truths and misleading leaps of logic, we still labeled them as misinformation tweets. For example, Tweet 2 in Table~\ref{tab:tweetannotation} seems to be spreading a questionable story about the installation of 5G antennas, so we annotated it as misinformation. If the tweet is judged to be countering a questionable story (see Tweet 5 in Table~\ref{tab:tweetannotation}), we annotated it as a counter-misinformation tweet. 


\begin{table*}[htbp]
\caption{Examples of tweets and assigned labels from our dataset.
}
\label{tab:tweetannotation}
\begin{center}
\begin{tabular}{|p{0.25cm}|p{7cm}|p{0.5cm}|p{1.65cm}|p{6.5cm}|} 
\toprule
 \textbf{No} & \textbf{Tweet} & \textbf{Topic}& \textbf{Label} & \textbf{Reason}\\  
\midrule
1 & Garlic onion and ginger are very nutritious, which has great potential to prevent corona virus. & Fake cures & Misinformation & Misinformation probability = 0.79; counter-misinformation probability = 0.01; and irrelevant probability = 0.06. \\ \hline
2 & Corona virus is a way for corporations to install 5G without us being around \#theory \#coronavirus & 5G & Misinformation & Misinformation probability = 0.63; counter-misinformation probability = 0.12; and irrelevant probability = 0.03. \\ \hline

3 & A conspiracy theory falsely linking 5G to the coronavirus is getting traction on social media.…https://www.politifact.com/factchecks/2020/apr/03... 
& 5G & Professional fact-check & The tweet contains the link mentioned in the IFCN dataset. \\ \hline

4 & @X Major Newspaper had to put this out... Garlic: No effect on coronavirus. Sesame oil on the body: No effect on coronavirus. Herbal remedies: No effect on coronavirus.  Smoking: No effect on coronavirus. & Fake cures & Opinion-based Citizen Response & Counter-misinformation probability = 0.69; misinformation probability = 0.27; irrelevant probability = 0.07; and no external link. \\ \hline

5 & how do some people seriously believe that coronavirus is not a virus and it’s the government trying to kill us with 5g network & 5G & Opinion-based Citizen Response & Counter-misinformation probability = 0.77; misinformation probability = 0.13; irrelevant probability = 0.06; and no external link. \\ \hline


6 & Garlic won’t keep the coronavirus at bay. Neither will saltwater gargling or cow dung https://www.scmp.com/week-asia/health-environment/article/3049261/garlic-cant-keep...
& Fake cures & Evidence-based Citizen Response & Counter-misinformation probability = 0.97; misinformation probability = 0.06; irrelevant probability = 0.01 and has an external link with evidence. \\ \hline

7 & \#Indonesia is grappling with surging garlic prices as the fast-spreading COVID spurs fears over supply disrupt in \#China. & Fake cures & Irrelevant & Irrelevant probability = 0.96; misinformation probability = 0.02; and counter-misinformation probability = 0.06. \\ 

\bottomrule
\end{tabular}
\end{center}
\end{table*}

\section{Misinformation and Counter-Misinformation Tweet Classifier}

In this section, we describe the text-based classifier that we developed to classify the tweets in our \covid\ tweet dataset as misinformation, counter-misinformation, or irrelevant tweets. Specifically, we first build the tweet representation for each tweet and then use the hand-labeled tweets to train the text-based classifier. 

\subsection{Tweet Representation}

\subsubsection{Embedding-based Representation}
Due to the advanced development in natural language processing field, we embed each tweet by the popular BERT text-embedding model~\cite{devlin2018bert}) to capture the semantic meaning of tweets. We selected BERT embedding because it obtained better performance than other models in our preliminary testing (e.g., GloVe~\cite{pennington2014glove}). We first remove all Twitter- and web-specific content such as URLs, usernames, hashtags and emojis, and then feed the remaining tweet text into the BERT model. After this processing, we select the final hidden layer result in BERT model as the resulting 768-dimensional tweet representation.

\subsubsection{Hashtag-based Representation}
Different from the embedding-based approach, in the tweet text, since people usually use hashtags (e.g., \#5G) to highlight important information, we used hashtags as information indicators to represent tweets. The hashtag-based representation counts the number of occurrences of each hashtag in the tweet text and selects the widely-mentioned hashtags in the dataset. In total, 768 hashtags were selected by popularity (i.e., their number of occurrence), keeping the same 768 dimension as that of the aforementioned embedding-based representation to have a fair comparison. We also varied the number of hashtags used (from 100 to 1,000), which achieved a comparable performance.

\subsection{Classifier Creation}
Given this standard three-class classification task, i.e., misinformation vs. counter-misinformation vs. irrelevant tweets, we trained three separate one-vs-all Logistic Regression classifiers. Note that we can also plug in any other classification model, such as SVM. Our empirical testing showed that the results of other models such as SVM were similar to the logistic regression results. Due to the relatively small dataset, we utilized the embedding-based BERT model rather than the fine-tuning-based BERT model or other advanced deep neural network models~\cite{devlin2018bert}) to avoid the possible over fitting problem. Each one-vs-all classifier is trained with hashtag-based, embedding-based and hashtag+embedding features, where the two feature vectors were concatenated. We conducted five-fold cross-validation on the manually-annotated tweets. The performance of the models was measured using precision, recall and F1 scores (see Table~\ref{tab:feature-test} and Table~\ref{tab:feature-test-fake-cure}). We report these scores together with the corresponding standard deviations.



\begin{table}[htbp]
\centering
\caption{Tweet classification performance for 5G topic.}
\resizebox{\columnwidth}{!}{%

\begin{tabular}{llll}
\toprule
                                           Feature &   Precision ($\sigma$) &      Recall ($\sigma$) &    F1 score ($\sigma$) \\
\midrule
\textbf{Misinformation Tweet Detection} &             &             &             \\
                         Hashtag&  0.386±0.01 &  0.859±0.02 &  0.533±0.01 \\
                           Embedding&  \textbf{0.763}±0.03 &  \textbf{0.788}±0.02 &  \textbf{0.775}±0.02 \\
                 Hashtag+Embedding&  0.758±0.02 &  0.782±0.02 &  0.770±0.02 \\
                 \hline
\textbf{Counter-Misinformation Tweet Detection} &             &             &             \\
                         Hashtag&  0.626±0.06 &  0.220±0.04 &  0.325±0.05 \\
                           Embedding&  \textbf{0.807}±0.05 &  \textbf{0.823}±0.01 &  \textbf{0.814}±0.02 \\
                 Hashtag+Embedding&  0.800±0.05 &  0.816±0.01 &  0.807±0.02 \\
                 \hline
\textbf{Irrelevant Tweet Detection} &             &             &             \\
                         Hashtag&  0.608±0.03 &  0.266±0.02 &  0.369±0.02 \\
                           Embedding&  \textbf{0.845}±0.02 &  \textbf{0.795}±0.05 &  \textbf{0.818}±0.03 \\
                 Hashtag+Embedding&  0.834±0.02 &  0.786±0.05 &  0.809±0.03 \\
\bottomrule
\end{tabular}
}
\label{tab:feature-test}
\end{table}

\begin{table}[htbp]
\centering
\caption{Tweet classification performance for fake cures topic.}
\resizebox{\columnwidth}{!}{
\begin{tabular}{llll}
\toprule
                                           Feature &   Precision ($\sigma$) &      Recall ($\sigma$) &    F1 score ($\sigma$) \\
\midrule
\textbf{Misinformation Tweet Detection} &             &             &             \\
                         Hashtag&  0.381±0.01 &  0.785±0.05 &  0.513±0.02 \\
                           Embedding&  \textbf{0.698}±0.03 &  \textbf{0.693}±0.04 &  \textbf{0.694}±0.03 \\
                 Hashtag+Embedding&  0.691±0.02 &  0.687±0.04 &  0.688±0.02 \\
                 \hline
\textbf{Counter-Misinformation Tweet Detection} &             &             &             \\
                         Hashtag&  0.563±0.09 &  0.178±0.02 &  0.270±0.02 \\
                           Embedding&  \textbf{0.760}±0.05 &  \textbf{0.759}±0.05 &  \textbf{0.758}±0.03 \\
                 Hashtag+Embedding&  0.748±0.05 &  0.744±0.04 &  0.744±0.03 \\
                 \hline
\textbf{Irrelevant Tweet Detection} &             &             &             \\
                         Hashtag&  0.475±0.02 &  0.295±0.03 &  0.363±0.03 \\
                           Embedding&  \textbf{0.677}±0.03 &  \textbf{0.678}±0.05 &  \textbf{0.676}±0.03 \\
                 Hashtag+Embedding&  0.675±0.03 &  0.678±0.04 &  0.676±0.03 \\
\bottomrule
\end{tabular}
}
\label{tab:feature-test-fake-cure}
\end{table}

As observed from Table~\ref{tab:feature-test} and Table~\ref{tab:feature-test-fake-cure}, by using the tweet embedding feature, we achieved the best performance with respect to precision, recall and F1 score. 
We noticed that the decreased performance in the combination of hashtag and embedding representations was achieved when unrelated features were added to the tweets. Therefore, we used embedding feature together with logistic regression classifier to label all tweets. To achieve high-confidence labelling results, we set a very strict rule, which states that if a tweet is classified as misinformation tweet, the score needs to be higher than 0.5 and  less than 0.5 for the other models (i.e., countering-misinformation and irrelevant). We used the same rule to detect counter-misinformation and irrelevant tweets. We use the 0.5 threshold because this value is widely-used in binary classification tasks~\cite{freeman2008comparison}.



Applying the classifier as described above returns a dataset that contains 33,237 misinformation tweets, 33,413 counter-misinformation tweets, and 441 professional fact checking tweets (see Table~\ref{tab:data-statistics}). In the following sections we further analyze these three categories of tweets.

\begin{table}[htbp]
\centering
\caption{Twitter Data Statistics}
\resizebox{\columnwidth}{!}{%
\begin{tabular}{l|r|r|r}
\toprule
 & 5G & Fake Cures & Total \\
\midrule
Total number of tweets & 97,283 & 58,185 & 155,468  \\
Number of Misinformation Tweets & 24,606   &  8,631 &  33,237 \\
Number of Professional Fact-check Tweets &  272 &  169 & 441 \\
Number of Citizen Counter-Misinformation Tweets & 26,229  &  8,184 &  33,413 \\
Number of Irrelevant Tweets & 46,176 &  41,201 &  87,377 \\
\bottomrule
\end{tabular}
}
\label{tab:data-statistics}
\end{table}

\section{Types of Counter-Misinformation Tweets}

One of the aims of this work is to understand whether citizens use professional fact-checks as evidence for countering misinformation. 
We find that 67\% of citizen responses do not contain any evidence to refute misinformation and we refer to these tweets as \textit{`opinion-based citizen responses'}.
On the other hand, only 33\% of citizen response tweets have URLs, out of which only 4\% point to fact-checking websites, showing the low use of the work done by professional fact checkers. To check where these URLs refer to, we randomly sampled 300 of these tweets and manually verified their URL. We found that 204 of them provided reliable evidence, of which 96 contained links to high-credibility sources, and the rest contained links to articles, videos, or websites that directly refute the misinformation. 
Similar to Chen et al.~\cite{chen2020neutral}, we defined high-credibility and low-credibility sources by compiling a list from several recent research papers~\cite{Shao2018anatomy,avram2020exposure}.
Since the proportion of links pointing to reliable evidence is very high, we label all counter-misinformation tweets with URLs as \emph{`evidence-based citizen responses'}. 





\begin{figure}[htbp]
\centering
  \includegraphics[scale=0.155]{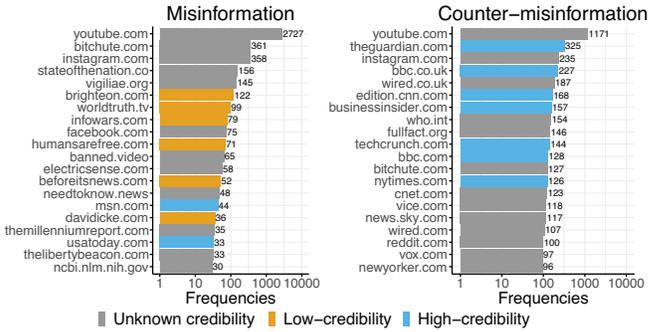}
  \caption{Most frequent external sources. High-credibility sources are more common in counter-misinformation tweets.}
  \label{fig:mostfrequentdomains}
\end{figure}

The frequency of high-credibility sources in counter-misinformation citizen responses is evidenced in Figure~\ref{fig:mostfrequentdomains}, which shows that 7 of the Top 20 most frequent sources are known high credibility. With respect to misinformation, we find that 5 of the Top 20 most-frequent sources are listed as low-credibility (see Figure~\ref{fig:mostfrequentdomains}). This finding confirms previous work which showed a strong correlation between misinformation and low-credibility sources~\cite{Shao2018anatomy}. 

In summary, in this section, we find that while concerned citizens use URLs to refute misinformation often, professional fact-checks are rarely used as evidence.
Research needs to investigate techniques that would make fact-checks more accessible to concerned citizens.

\section{Characterizing the spread of Counter-Misinformation}

To get a better understanding of the extent to which misinformation is being countered, we investigate tweet volume, engagement, and spread. 

\begin{figure*}[htbp]
\centering
  \includegraphics[scale=0.23]{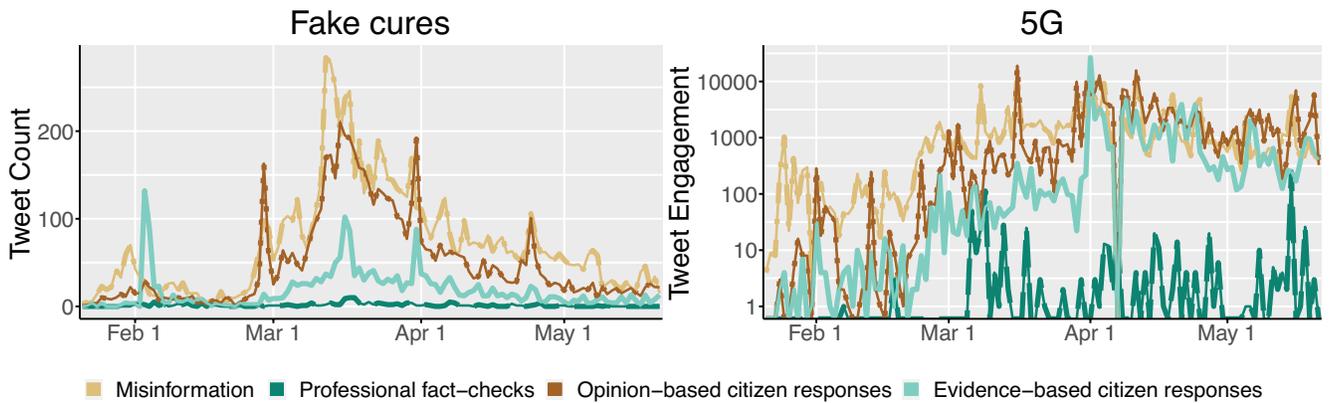}
  \caption{Tweet volume and engagement over time. Both volume and engagement of countering misinformation are comparable to misinformation. 
  }
  \label{fig:volumeandengagement}
\end{figure*}

\subsection{Misinformation has higher volume} 
Figure~\ref{fig:volumeandengagement} (left) shows the volume in terms of number of tweets of fake cures topic over the chosen four-month period. To provide an accurate representation of tweet volume and engagement, we selected unique tweets and discarded retweets. 
First, we note that for both topics, the volume of professional fact-checks is always significantly lower than that of the other groups.
Second, we find that for 5G, the volume of misinformation tweets is comparable to the volume of counter-misinformation tweets (Wilcoxon pairwise comparison p $>$ 0.05). 
Next, we note that among concerned citizens, the volume of tweets that counter misinformation without using professional fact-checking URLs as evidence is several times higher than those that use evidence. 
These findings show that there is a comparable amount of people that are countering misinformation and that tweets by professional fact-checkers are still not popular across the twitter community. 

Finally, an important observation is that there is a strong positive correlation between misinformation and counter-misinformation tweet counts, both for fake cures (Spearman correlation $\rho$ = 0.88, p $<$ 0.001) and 5G ($\rho$ = 0.85, p $<$ 0.001). Importantly, for both topics, we found a lower, but still positive, correlation between misinformation and fact-checks (fake cures: $\rho$ = $0.45$, p $<$ 0.001, 5G: $\rho$ = 0.52, p $<$ 0.001). 
These findings show that misinformation is being countered at the same rate as it is being spread, similar to the findings made in Mendoza et al.~\cite{mendoza2010twitter}. 

\subsection{Counter-misinformation and misinformation attract similar tweet engagement} 

We investigate engagement volume because engaging with a tweet increases its visibility. Tweet engagement is defined as liking or favoriting the tweet and sharing a tweet with or without an additional comment. Figure~\ref{fig:volumeandengagement} (right) shows the engagement results for the 5G topic. We used Wilcoxon pairwise test to compare the differences across the types of information. First, we note that the engagement of misinformation and counter-misinformation tweets is similar (p $>$ 0.05). This means that concerned citizens are obtaining a comparable level of engagement as users that spread misinformation. This is not the case with professional fact-checkers because in all cases they obtained significantly lower amount of engagement than misinformation and counter-misinformation. This indicates that work is needed to increase the visibility of professional fact-checking efforts. Finally, when restricting our analysis to only retweets, we find that most of the tweets received 0 retweets (71\% - 80\%), while only few tweets were retweeted more than 100 times (0.2\% - 0.6\%). This finding shows that few counter-misinformation and misinformation tweets get propagated through the Twitter community.

Overall, in this section, we find that concerned citizens are producing a similar volume of tweets as misinformation spreaders. Similarly, engagement levels are comparable and most tweets do not receive more than 100 retweets. Moreover, for 5G, counter-misinformation tweets receive more engagement than misinformation tweets. Professional fact-checks receive a significantly lower volume of engagement than misinformation and concerned citizen tweets.

\section{Characterizing Textual Properties of Counter-Misinformation}
In this section, we analyze the linguistic and textual properties of misinformation, counter-misinformation, and fact-checks. 
We investigate their sentiment, politeness, word use and LIWC characteristics. 
This investigation will provide insights on how linguistic and textual properties could be used to detect and counter misinformation.

\subsection{Sentiment and politeness of citizen responses}

Here we analyze the sentiment and politeness~\cite{chang2020convokit} of counter-misinformation responses made by citizens. 
Prior research has also found it to be useful in detecting misinformation~\cite{ajao2019sentiment}.
Table~\ref{tab:politenesssentiment}
provides a few examples of tweets and their corresponding politeness and sentiment scores. 

\begin{figure}[htbp]
\centering
  \includegraphics[scale=0.17]{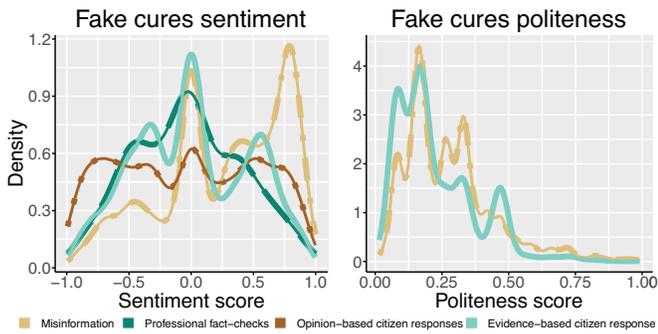}
  \caption{Sentiment and politeness for fake cures. Misinformation tweets are more positive and professional fact-checks are neutral (left). Evidence-based citizen responses are less polite than misinformation (right).
  }
  \label{fig:sentimentpoliteness}
\end{figure}

Figure~\ref{fig:sentimentpoliteness} compares the distribution of sentiment and politeness across different types of content. 
For sentiment, we find that tweets by professional fact-checkers are neutral. On the other hand, tweets spreading misinformation about fake cures of \covid\ are significantly positive, potentially hailing the `cure' that works like a miracle. However, counter-misinformation tweets tend to be more neutral. 

\begin{table}[htbp]
\caption{Examples of extreme levels of politeness and sentiment.}
\label{tab:politenesssentiment}
\begin{center}
\begin{tabular}{|p{6.8cm}|p{1cm}|} 
\hline
 \textbf{Tweet} & \textbf{Type}\\ \hline
@X,@Y @Z Thank you for the work that you do. Currently there has been a lot of misinformation surrounding the spread of Covid 19 through 5G. & Polite \\ \hline
Y’all idiots really talking about \#5g causes \#COVID19 .. lol Iran has no 5g towers hell 95\% of America doesn’t have 5g towers. No towers on cruise ship either but people are getting sick y’all have no answers & Impolite \\ \hline\hline
@X, @Y My uncle is a smart, handsome and professional expert. He told me that 5G does not cause coronavirus.
& Positive sentiment 
\\ \hline
If you think COVID-19 is actually a hoax and the sickness is from 5G radiation poisoning f**k you’re an idiot. DUMB. You read something on FB and think you’re a radiologist/scientist when really you’re just dumb dumb dumbbb & Negative sentiment\\ \hline

\end{tabular}
\end{center}
\end{table}

As seen in Figure~\ref{fig:sentimentpoliteness}, with regards to politeness, we find that misinformation tweets are significantly more polite than countering-misinformation tweets (p $<$ 0.001). Surprisingly, professional fact-checks are quantified as less polite as well. This is because these tweets do not contain much text, and the formal and restrained nature of these tweets could be leading to this outcome~\cite{maros2017politeness}. Previous research~\cite{brown1987politeness} defined polite text to be direct, clear, unambiguous and concise~\cite{brown1987politeness}. This means that misinformation tweets seem to be exhibiting more of these characteristics than the other tweets.  

Digging deeper into the citizen responses to fake cures misinformation, we find that evidence-based citizen responses are very similar to professional fact-checking tweets in terms of sentiment, most of them being neutral. On the other hand, opinion-based citizen responses tend to express more sentiment, exhibiting a uniform distribution through the spectrum of sentiment scores.

The main outcome of these findings is that \covid\ misinformation is mostly positive, this contrasts with previous research which showed that misinformation is mostly negative~\cite{jiang2018linguistic}. This finding is further evidence that misinformation spreaders are continuously changing their tactics~\cite{kumar2018false}. The lack of politeness, positive and negative sentiments in professional fact-checking tweets is also an interesting finding. This finding requires further research to investigate whether manipulating these properties could lead to higher volume and engagement.


\subsection{Psycholinguistic Characteristics of Citizen Responses}
We analyze textual properties using LIWC dictionary~\cite{pennebaker2001linguistic}, which accounts
for the psycholinguistic properties of the text.

\begin{figure*}[htbp]
\centering
  \includegraphics[scale=0.14]{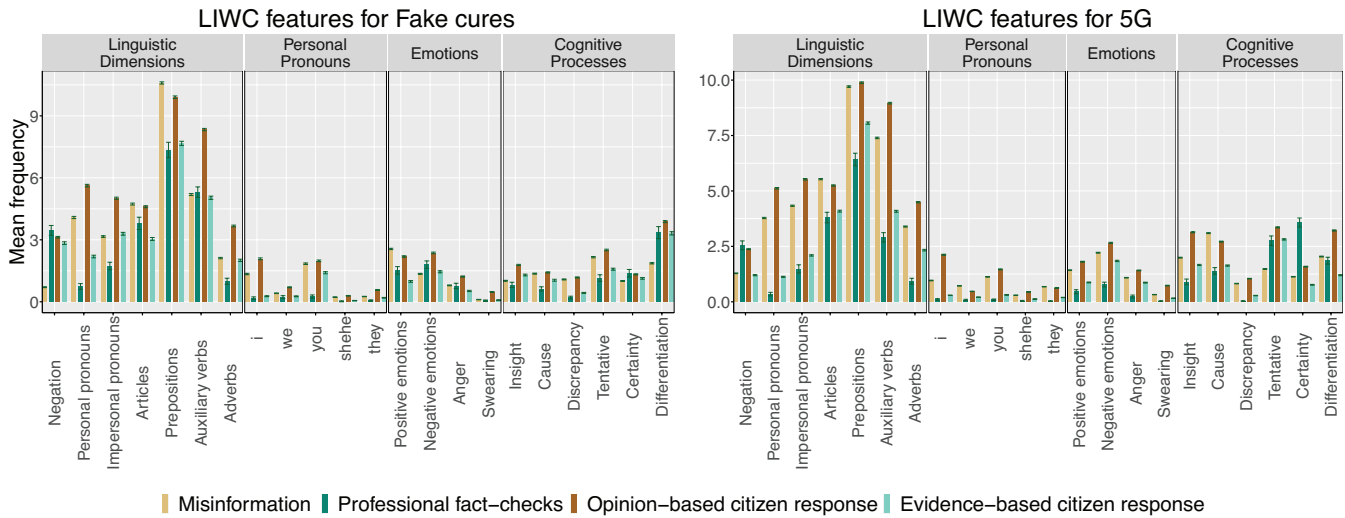}
  \caption{LIWC results for fake cures (left) and 5G (right) tweets. Opinion-based citizen responses have similar linguistic properties as misinformation.}
  \label{fig:liwcboth}
\end{figure*}

First, opinion-based citizen response tweets (see Figure~\ref{fig:liwcboth}) are significantly more self centered, use `you' and third person pronouns (i.e., `he' or 'she') more than misinformation and evidence-based responses (p $<$ 0.001).
We also find that opinion-based response tweets have a significantly higher amount of swear words than the other tweets (p $<$ 0.001).
Together, these two findings show that citizens are engaging in a more direct, confrontational, and argumentative conversation with misinformation spreaders. 
While swearing and confrontation used in countering misinformation contrasts with the findings from previous research~\cite{jiang2018linguistic}, our analysis shows that it is a common technique used by citizen responders to call out misinformation spreaders~\cite{memon2020characterizinga}.


Second, our analysis shows that evidence-based responses are quite similar to professional fact-checking tweets. For instance, we did not find significant differences when comparing all five pronouns (p $>$ 0.05). We find similar values for impersonal pronouns, authentic words, dictionary words, words greater than 6 letters and tone. They have similar amount of words that contain discrepancy, tentative, achievement, power, reward, assent, fillers, netspeak and non-fluence. Also, we find that professional fact-checkers infrequently use personal and impersonal pronouns, especially the `i' and `he'/`she'. 

Third, the textual properties of opinion-based citizen responses and misinformation tweets have highly similar levels of achievement, anxiety, positive emotions, netspeak and non-fluence (see Fig~\ref{fig:liwcboth}). Furthermore, they have similar amounts of adverbs, articles, auxiliary verbs, prepositions, dictionary words, function, long words, and tone.
Opinion-based citizen responses have significantly more insight words, differentiation words, and tentative words (p $<$ 0.001). On the other hand, misinformation tweets exhibit significantly more words related to cause (p $<$ 0.001). For fake cures, counter-misinformation without evidence show significantly higher affiliation and tweets that counter-misinformation with evidence show significantly higher risk (p $<$ 0.001). 



In summary, we find that opinion-based counter-misinformation portray insight, exhibit negative emotions and anger by using more negative words, including swear words. 
This implies that concerned citizens are engaging in a more direct, confrontational, and argumentative conversation and do not shy away from calling out misinformation spreaders~\cite{memon2020characterizinga}.

\section{User Characteristics of Concerned Citizens}

To have a better understanding of concerned citizens who respond to misinformation we investigate their accounts' age, their popularity and whether they have verified accounts.

\subsection{Misinformation spreaders have more recent accounts than concerned citizens}
First, we find that accounts that spread misinformation (fake cures: 5.7 years; 5G: 6 years) are significantly (p $<$ 0.001) more recent than concerned citizens accounts (fake cures: 6.5 years; 5G: 6.8 years). Second, we find that professional fact-checking accounts (fake cures: 22\%; 5G: 8\%) and concerned citizens accounts (fake cures: 7\% years; 5G: 4\%) are significantly more likely (p $<$ 0.001) to be verified accounts than misinformation (1\%). These findings show that accounts that spread misinformation are typically more recent~\cite{shahi2020exploratory} and concerned citizens are more likely to have verified accounts than misinformation spreaders.

\subsection{Concerned citizens are more popular}
Account popularity is a relevant user characteristic because when a user posts a tweet, it becomes visible to all of their followers. We find that users that spread misinformation have significantly (p $<$ 0.001) fewer followers (fake cures: 6,052 and 5G: 6,651) than concerned citizens (fake cures: 40,295; 5G: 22,133). Professionals that fact-checked fake cures had significantly more followers (= 85,586) than concerned citizens and misinformation spreaders. Contrarily, professionals that fact-checked 5G conspiracy theories had similar number of followers as misinformation spreaders (8,169 vs 6,651; p $>$ 0.05) and significantly fewer followers than concerned citizens (8,169 vs 22,133;  p $<$ 0.001). These findings show that counter-misinformation has the potential of reaching a wider audience than misinformation. Despite receiving low volume and engagement, some professional fact-checking accounts have a considerably large audience.

In summary, in this section, we find that concerned citizens are more popular and have older accounts, indicating that counter-misinformation has the potential to reach more users and are more trusted. These accounts are more likely to have verified accounts than misinformation spreaders.

\section{Conclusion}

In this work, we focus on \covid\ related misinformation on Twitter and utilize a data-driven approach to investigate how fact checks and other organic user responses attempt to refute such misinformation.
 
Overall, our work shows that countering of misinformation cannot be studied by only considering fact checks because 96\% of all refutations are being done by concerned citizens (i.e., the crowd). 
Moreover, professional fact-checks are not only low in tweet volume, but also receive a significantly lower number of retweets as opposed to those received by the crowd. Our analysis also reveals that the crowd uses two methods to counter-misinformation. The first approach is evidence-based, which includes links to fact-checks or other trusted sources. The second is opinion-based, which does not include checkable evidence. We find that opinion-based tweets include more assertive language, use more negative words, are more abusive, and exhibit negative emotions and anger.

Our results reveal interesting insights and can potentially lead to a better understanding of how to leverage the organic countering of misinformation to support and amplify fact-checking efforts best.
More importantly this work could also lead to development of tools and mechanisms that can empower concerned citizens to combat misinformation.

Finally, the work presented has a number of limitations that could be addressed in the future. First, we only focus on the Twitter platform. In future work, one can extend our analysis to different platforms, like WhatsApp, Facebook and Instagram, since previous research found that misinformation flows across multiple platforms~\cite{cinelli2020covid}. Second, we only studied two misinformation topics. These two topics help us understand the role of fact checks and organic citizens, but there could be variations across areas that are targets of misinformation (e.g., elections and geopolitical issues). 
In fact, we did find evidence of differences across topics which needs to be fully explored in the future.

\section*{Acknowledgements}
This research has been supported in part by NSF IIS2027689, New York University Abu Dhabi, Georgia Tech IDEaS, Adobe, and Microsoft Azure.

\bibliographystyle{IEEEtran}
\bibliography{IEEEabrv,main}

\begin{thebibliography}{10}
\providecommand{\url}[1]{#1}
\csname url@samestyle\endcsname
\providecommand{\newblock}{\relax}
\providecommand{\bibinfo}[2]{#2}
\providecommand{\BIBentrySTDinterwordspacing}{\spaceskip=0pt\relax}
\providecommand{\BIBentryALTinterwordstretchfactor}{4}
\providecommand{\BIBentryALTinterwordspacing}{\spaceskip=\fontdimen2\font plus
\BIBentryALTinterwordstretchfactor\fontdimen3\font minus
  \fontdimen4\font\relax}
\providecommand{\BIBforeignlanguage}[2]{{%
\expandafter\ifx\csname l@#1\endcsname\relax
\typeout{** WARNING: IEEEtran.bst: No hyphenation pattern has been}%
\typeout{** loaded for the language `#1'. Using the pattern for}%
\typeout{** the default language instead.}%
\else
\language=\csname l@#1\endcsname
\fi
#2}}
\providecommand{\BIBdecl}{\relax}
\BIBdecl

\bibitem{who2020novel}
\BIBentryALTinterwordspacing
WHO, \emph{Novel {C}oronavirus (2019-nCoV)}, 2020 (accessed August 26, 2020).
  [Online]. Available:
  \url{https://www.who.int/docs/default-source/coronaviruse/situation-reports/20200202-sitrep-13-ncov-v3.pdf}
\BIBentrySTDinterwordspacing

\bibitem{mendoza2010twitter}
M.~Mendoza, B.~Poblete, and C.~Castillo, ``Twitter under crisis: Can we trust
  what we rt?'' in \emph{Proceedings of the first workshop on social media
  analytics}, 2010, pp. 71--79.

\bibitem{viewing2010microblogging}
S.~Vieweg, A.~L. Hughes, K.~Starbird, and L.~Palen, ``Microblogging during two
  natural hazards events: What twitter may contribute to situational
  awareness,'' in \emph{Proceedings of the SIGCHI Conference on Human Factors
  in Computing Systems}, ser. CHI ’10.\hskip 1em plus 0.5em minus 0.4em\relax
  New York, NY, USA: Association for Computing Machinery, 2010, p. 1079–1088.

\bibitem{burnap2014tweeting}
P.~Burnap, M.~L. Williams, L.~Sloan, O.~Rana, W.~Housley, A.~Edwards,
  V.~Knight, R.~Procter, and A.~Voss, ``Tweeting the terror: modelling the
  social media reaction to the woolwich terrorist attack,'' \emph{Social
  Network Analysis and Mining}, vol.~4, no.~1, p. 206, 2014.

\bibitem{starbird2014rumors}
K.~Starbird, J.~Maddock, M.~Orand, P.~Achterman, and R.~M. Mason, ``Rumors,
  false flags, and digital vigilantes: Misinformation on twitter after the 2013
  boston marathon bombing,'' \emph{IConf. 2014 Proc.}, 2014.

\bibitem{oyeyemi2014ebola}
S.~O. Oyeyemi, E.~Gabarron, and R.~Wynn, ``Ebola, twitter, and misinformation:
  a dangerous combination?'' \emph{BMJ}, vol. 349, 2014.

\bibitem{miller2017what}
M.~Miller, T.~Banerjee, R.~Muppalla, W.~Romine, and A.~Sheth, ``What are people
  tweeting about zika? an exploratory study concerning its symptoms, treatment,
  transmission, and prevention,'' \emph{JMIR Public Health Surveill}, vol.~3,
  no.~2, p. e38, Jun 2017.

\bibitem{ortiz2017yellow}
Y.~Ortiz-Mart{\'\i}nez and L.~F. Jim{\'e}nez-Arcia, ``Yellow fever outbreaks
  and twitter: Rumors and misinformation,'' \emph{American Journal of Infection
  Control}, vol.~45, no.~7, pp. 816--817, 2017.

\bibitem{bbc2020novel}
\BIBentryALTinterwordspacing
BBC, \emph{Coronavirus: `Murder threats` to telecoms engineers over 5{G}}, 2020
  (accessed August 26, 2020). [Online]. Available:
  \url{https://www.bbc.com/news/newsbeat-52395771}
\BIBentrySTDinterwordspacing

\bibitem{islam2020covid-19}
M.~S. Islam, T.~Sarkar, S.~H. Khan, A.-H. Mostofa~Kamal, S.~M.~M. Hasan,
  A.~Kabir, D.~Yeasmin, M.~A. Islam, K.~I. Amin~Chowdhury, K.~S. Anwar, A.~A.
  Chughtai, and H.~Seale, ``Covid-19-related infodemic and its impact on public
  health: A global social media analysis,'' \emph{The American Journal of
  Tropical Medicine and Hygiene}, 2020.

\bibitem{wintersieck2017debating}
A.~L. Wintersieck, ``Debating the truth: The impact of fact-checking during
  electoral debates,'' \emph{American Politics Research}, vol.~45, no.~2, pp.
  304--331, 2017.

\bibitem{shahi2020exploratory}
G.~K. Shahi, A.~Dirkson, and T.~A. Majchrzak, ``An exploratory study of
  covid-19 misinformation on twitter,'' \emph{arXiv:2005.05710}, 2020.

\bibitem{lazer2018thescience}
D.~M.~J. Lazer, M.~A. Baum, Y.~Benkler, A.~J. Berinsky, K.~M. Greenhill,
  F.~Menczer, M.~J. Metzger, B.~Nyhan, G.~Pennycook, D.~Rothschild,
  M.~Schudson, S.~A. Sloman, C.~R. Sunstein, E.~A. Thorson, D.~J. Watts, and
  J.~L. Zittrain, ``The science of fake news,'' \emph{Science}, vol. 359, no.
  6380, pp. 1094--1096, 2018.

\bibitem{kumar2018false}
S.~Kumar and N.~Shah, ``False information on web and social media: A survey,''
  \emph{arXiv preprint arXiv:1804.08559}, 2018.

\bibitem{brennen2020types}
J.~S. Brennen, F.~Simon, P.~N. Howard, and R.~K. Nielsen, ``Types, sources, and
  claims of covid-19 misinformation,'' \emph{Reuters Institute}, vol.~7, 2020.

\bibitem{wang2020weak}
Y.~Wang, W.~Yang, F.~Ma, J.~Xu, B.~Zhong, Q.~Deng, and J.~Gao, ``Weak
  supervision for fake news detection via reinforcement learning,'' in
  \emph{Proceedings of the AAAI Conference on Artificial Intelligence},
  vol.~34, no.~01, 2020, pp. 516--523.

\bibitem{wang2018eann}
Y.~Wang, F.~Ma, Z.~Jin, Y.~Yuan, G.~Xun, K.~Jha, L.~Su, and J.~Gao, ``Eann:
  Event adversarial neural networks for multi-modal fake news detection,'' in
  \emph{Proceedings of the 24th acm sigkdd international conference on
  knowledge discovery \& data mining}, 2018, pp. 849--857.

\bibitem{young2020detecting}
S.~Young, ``Detecting covid-19 misinformation on social media,'' 2020.

\bibitem{friggeri2014rumor}
A.~Friggeri, L.~Adamic, D.~Eckles, and J.~Cheng, ``Rumor cascades,'' in
  \emph{8th International AAAI Conference on Weblogs and Social Media}, 2014.

\bibitem{vosoughi2018spread}
S.~Vosoughi, D.~Roy, and S.~Aral, ``The spread of true and false news online,''
  \emph{Science}, vol. 359, no. 6380, pp. 1146--1151, 2018.

\bibitem{vraga2017using}
E.~K. Vraga and L.~Bode, ``Using expert sources to correct health
  misinformation in social media,'' \emph{Science Communication}, vol.~39,
  no.~5, pp. 621--645, 2017.

\bibitem{van2020seeking}
T.~G. van~der Meer and Y.~Jin, ``Seeking formula for misinformation treatment
  in public health crises: The effects of corrective information type and
  source,'' \emph{Health Communication}, vol.~35, pp. 560--575, 2020.

\bibitem{dharawat2020drink}
A.~R. Dharawat, I.~Lourentzou, A.~Morales, and C.~Zhai, ``Drink bleach or do
  what now? covid-hera: A dataset for risk-informed health decision making in
  the presence of covid19 misinformation,'' 2020.

\bibitem{kumar2016disinformation}
S.~Kumar, R.~West, and J.~Leskovec, ``Disinformation on the web: Impact,
  characteristics, and detection of wikipedia hoaxes,'' in \emph{Proceedings of
  the 25th international conference on World Wide Web}, 2016, pp. 591--602.

\bibitem{hassan2019introduction}
N.~Hassan, C.~Li, J.~Yang, and C.~Yu, ``Introduction to the special issue on
  combating digital misinformation and disinformation,'' 2019.

\bibitem{antoniadis2015model}
S.~Antoniadis, I.~Litou, and V.~Kalogeraki, ``A model for identifying
  misinformation in online social networks,'' in \emph{OTM Confederated
  International Conferences" On the Move to Meaningful Internet
  Systems"}.\hskip 1em plus 0.5em minus 0.4em\relax Springer, 2015, pp.
  473--482.

\bibitem{litou2017efficient}
I.~Litou, V.~Kalogeraki, I.~Katakis, and D.~Gunopulos, ``Efficient and timely
  misinformation blocking under varying cost constraints,'' \emph{Online Social
  Networks and Media}, vol.~2, pp. 19--31, 2017.

\bibitem{bode2020right}
L.~Bode, E.~K. Vraga, and M.~Tully, ``Do the right thing: tone may not affect
  correction of misinformation on social media,'' \emph{Harvard Kennedy School
  Misinformation Review}, 2020.

\bibitem{bhuiyan2020investigating}
M.~M. Bhuiyan, A.~X. Zhang, C.~M. Sehat, and T.~Mitra, ``Investigating
  “who” in the crowdsourcing of news credibility,'' in \emph{Computational
  Journalism Symposium}, 2020.

\bibitem{roitero2020covid}
K.~Roitero, M.~Soprano, B.~Portelli, D.~Spina, V.~Della~Mea, G.~Serra,
  S.~Mizzaro, and G.~Demartini, ``The covid-19 infodemic: Can the crowd judge
  recent misinformation objectively?'' in \emph{Proceedings of the 29th ACM
  International Conference on Information \& Knowledge Management}, 2020, pp.
  1305--1314.

\bibitem{cinelli2020covid}
M.~Cinelli, W.~Quattrociocchi, A.~Galeazzi, C.~M. Valensise, E.~Brugnoli, A.~L.
  Schmidt, P.~Zola, F.~Zollo, and A.~Scala, ``The covid-19 social media
  infodemic,'' \emph{arXiv preprint arXiv:2003.05004}, 2020.

\bibitem{kouzy2020coronavirus}
R.~Kouzy, J.~Abi~Jaoude, A.~Kraitem, M.~B. El~Alam, B.~Karam, E.~Adib,
  J.~Zarka, C.~Traboulsi, E.~W. Akl, and K.~Baddour, ``Coronavirus goes viral:
  quantifying the covid-19 misinformation epidemic on twitter,'' \emph{Cureus},
  vol.~12, no.~3, 2020.

\bibitem{sharma2020covid}
K.~Sharma, S.~Seo, C.~Meng, S.~Rambhatla, and Y.~Liu, ``Covid-19 on social
  media: Analyzing misinformation in twitter conversations,'' \emph{arXiv
  preprint arXiv:2003.12309}, 2020.

\bibitem{huang2020disinformation}
B.~Huang and K.~M. Carley, ``Disinformation and misinformation on twitter
  during the novel coronavirus outbreak,'' \emph{arXiv preprint
  arXiv:2006.04278}, 2020.

\bibitem{alam2020fighting}
F.~Alam, S.~Shaar, A.~Nikolov, H.~Mubarak, G.~D.~S. Martino, A.~Abdelali,
  F.~Dalvi, N.~Durrani, H.~Sajjad, K.~Darwish \emph{et~al.}, ``Fighting the
  covid-19 infodemic: Modeling the perspective of journalists, fact-checkers,
  social media platforms, policy makers, and the society,'' \emph{arXiv
  preprint arXiv:2005.00033}, 2020.

\bibitem{memon2020characterizinga}
S.~A. Memon and K.~M. Carley, ``Characterizing covid-19 misinformation
  communities using a novel twitter dataset,'' \emph{arXiv preprint
  arXiv:2008.00791}, 2020.

\bibitem{chen2020tracking}
E.~Chen, K.~Lerman, and E.~Ferrara, ``Tracking social media discourse about the
  covid-19 pandemic: Development of a public coronavirus twitter data set,''
  \emph{JMIR Public Health Surveill}, vol.~6, no.~2, May 2020.

\bibitem{hossain2020detecting}
T.~Hossain, R.~L. Logan~IV, A.~Ugarte, Y.~Matsubara, S.~Singh, and S.~Young,
  ``Detecting covid-19 misinformation on social media,'' 2020.

\bibitem{schuster2004note}
C.~Schuster, ``A note on the interpretation of weighted kappa and its relations
  to other rater agreement statistics for metric scales,'' \emph{Educational
  and Psychological Measurement}, vol.~64, no.~2, pp. 243--253, 2004.

\bibitem{zubiaga2015towards}
A.~Zubiaga, M.~Liakata, R.~Procter, K.~Bontcheva, and P.~Tolmie, ``Towards
  detecting rumours in social media,'' in \emph{Workshops at the Twenty-Ninth
  AAAI Conference on Artificial Intelligence}, 2015.

\bibitem{qazvinian2011rumor}
V.~Qazvinian, E.~Rosengren, D.~Radev, and Q.~Mei, ``Rumor has it: Identifying
  misinformation in microblogs,'' in \emph{Proceedings of the 2011 Conference
  on Empirical Methods in Natural Language Processing}, 2011, pp. 1589--1599.

\bibitem{devlin2018bert}
J.~Devlin, M.-W. Chang, K.~Lee, and K.~Toutanova, ``Bert: Pre-training of deep
  bidirectional transformers for language understanding,'' \emph{arXiv preprint
  arXiv:1810.04805}, 2018.

\bibitem{pennington2014glove}
J.~Pennington, R.~Socher, and C.~D. Manning, ``Glove: Global vectors for word
  representation,'' in \emph{Proc. of the 2014 conference on empirical methods
  in natural language processing}, 2014, pp. 1532--1543.

\bibitem{freeman2008comparison}
E.~A. Freeman and G.~G. Moisen, ``A comparison of the performance of threshold
  criteria for binary classification in terms of predicted prevalence and
  kappa,'' \emph{Ecological modelling}, vol. 217, no. 1-2, pp. 48--58, 2008.

\bibitem{chen2020neutral}
W.~Chen, D.~Pacheco, K.-C. Yang, and F.~Menczer, ``Neutral bots reveal
  political bias on social media,'' \emph{arXiv preprint arXiv:2005.08141},
  2020.

\bibitem{Shao2018anatomy}
C.~Shao, P.-M. Hui, L.~Wang, X.~Jiang, A.~Flammini, F.~Menczer, and G.~L.
  Ciampaglia, ``Anatomy of an online misinformation network,'' \emph{PLoS ONE},
  vol.~13, no.~4, p. e0196087, 2018.

\bibitem{avram2020exposure}
M.~Avram, N.~Micallef, S.~Patil, and F.~Menczer, ``{Exposure to social
  engagement metrics increases vulnerability to misinformation},''
  \emph{Harvard Kennedy School Misinformation Review}, jul 2020.

\bibitem{chang2020convokit}
J.~P. Chang, C.~Chiam, L.~Fu, A.~Z. Wang, J.~Zhang, and
  C.~Danescu-Niculescu-Mizil, ``Convokit: A toolkit for the analysis of
  conversations,'' \emph{arXiv preprint arXiv:2005.04246}, 2020.

\bibitem{ajao2019sentiment}
O.~{Ajao}, D.~{Bhowmik}, and S.~{Zargari}, ``Sentiment aware fake news
  detection on online social networks,'' in \emph{IEEE International Conference
  on Acoustics, Speech and Signal Processing}, 2019, pp. 2507--2511.

\bibitem{maros2017politeness}
M.~Maros and L.~Rosli, ``Politeness strategies in twitter updates of female
  english language studies malaysian undergraduates,'' \emph{3L: Language,
  Linguistics, Literature{\textregistered}}, vol.~23, no.~1, 2017.

\bibitem{brown1987politeness}
P.~Brown, S.~C. Levinson, and S.~C. Levinson, \emph{Politeness: Some universals
  in language usage}.\hskip 1em plus 0.5em minus 0.4em\relax Cambridge
  university press, 1987, vol.~4.

\bibitem{jiang2018linguistic}
S.~Jiang and C.~Wilson, ``Linguistic signals under misinformation and
  fact-checking: Evidence from user comments on social media,'' \emph{Proc. ACM
  Hum.-Comput. Interact.}, vol.~2, no. CSCW, Nov. 2018.

\bibitem{pennebaker2001linguistic}
J.~W. Pennebaker, M.~E. Francis, and R.~J. Booth, ``Linguistic inquiry and word
  count: Liwc 2001,'' \emph{Mahway: Lawrence Erlbaum Associates}, vol.~71, no.
  2001, p. 2001, 2001.

\end{thebibliography}

\end{document}